\documentstyle[twocolumn,aps,epsfig]{revtex}
\begin{document}
\topmargin .1cm
\draft
\title{Lattice spacing dependence of phase transition temperature 
  in the classical linear sigma model}
\author{\bf A. K. Chaudhuri\cite{byline}}
\address{ Variable Energy Cyclotron Centre\\
1/AF,Bidhan Nagar, Calcutta - 700 064\\}
\maketitle

\begin{abstract}

We have investigated the phase transition properties of classical
linear  sigma  model. The fields were kept in contact with a heat
bath for sufficiently long time such that fields are equilibrated
at the temperature of the heat bath. It was shown that the  sigma
model  fields  undergoes  phase  transition,  but  the transition
temperature depend crucially  on  the  lattice  spacing.  In  the
continuum  limit,  the  transition  temperature  tends to zero or
at least to a very low value.
\end{abstract}


The possibility of forming disoriented chiral condensate (DCC) in
relativistic  heavy  ion  collisions  has  generated considerable
research  activities  in  recent   years.   Anslem   and   Ryskin
\cite{an88}   first   considered  the  possibility  of  producing
classical pion fields in heavy ion collisions.  The  idea  became
popular after Rajagopal and Wilczek \cite{ra93}   proposed  the
quench  scenario and Bjorken et al \cite{bj93} proposed the Baked-Alaska
model for producing DCC. Rajagopal and Wilczek argued that for  a
second  order  chiral phase transition, the chiral condensate can
become temporarily disoriented in the  nonequilibrium  conditions
encountered  in  heavy  ion  collisions. As the temperature drops
below $T_c$, the chiral symmetry begins to  break  by  developing
domains  in  which  the  chiral field is misaligned from its true
vacuum value.  The  misaligned  condensate  has  the  same  quark
content   and   quantum  numbers  as  do  pions  and  essentially
constitute a  classical  pion  field.  The  system  will  finally
relaxes  to  the true vacuum and in the process can emit coherent
pions. Since the disoriented domains have  well  defined  isospin
orientation, the associated pions can exhibit novel centauro-like
\cite{la80} fluctuations of neutral and charged pions.

Most dynamical studies of DCC have been based on the linear sigma
model
\cite{bl92,as94,ga94a,ga94b,hu94,ra96,ra97,bi97,ri98,cs99,ch99a,ch99b}
in which the chiral degrees of freedom are described by the  real
O(4) field $\Phi=(\sigma,\roarrow{\pi})$, with the Lagrangian,

\begin{equation} {\mathcal{L}} =\frac{1}{2} (\partial_\mu \Phi)^2
- \frac{\lambda}{4} (\Phi^2 - f_\pi^2)^2, \end{equation}

\noindent  where  $\lambda$  is  a positive coupling constant and
$f_\pi$ is the pion decay constant. In the vacuum the symmetry is
spontaneously broken and the sigma  field  acquires  a  non  zero
vacuum  expectation  value  $<0|\sigma|0>  =  f_\pi$. Symmetry is
restored  by  a  second  order   phase   transition.   Transition
temperature   can   be   calculated   in  mean  field  level  as,
$T_c^2=12f_\pi^2/(N+2)$, with  N=4  \cite{ka96}.  In  this  model
pions  are  massless.  To  be  realistic,  one introduces a small
symmetry breaking term $H_\sigma$  to  the  lagrangian  and  pion
become  massive.  The  parameters  of  the  model can be fixed by
defining the meson masses and pion decay constant.

It  is  well  known that classical fields suffers from divergence
problem. Thermodynamic of classical fields are only defined if an
ultraviolet cut off  is  imposed  on  the  momentum.  In  lattice
simulations,  finite  lattice  spacing provides the cut off. Some
thermodynamic quantities  such  as  the  diffusion  rate  of  the
topological   charge   \cite{am91,am95}   in   non-abelian  gauge
theories,  or  the  Lyapunov  exponent   \cite{mu92}   which   is
equivalent  to  the  damping rate of soft thermal excitations are
found to be insensitive to the cut off. However, it is not known,
whether the critical temperature of phase transition  the  linear
sigma  model  depend  on  the  momentum  cut off or not. With the
possibility of experiments being performed at RHIC in  search  of
Quark-gluon Plasma, where one expects to find signal of DCC, this
investigation becomes important.

Aim of the present letter is to show that critical temperature of
linear sigma model do indeed depend on the lattice spacing or the
momentum cut off and in the continuum limit, the phase transition
temperature approaches zero. To see phase transition we have used
the  old  idea of finite temperature field theory, that is if you
attach  the  fields  with  a  heat   bath,   their   equilibrium
configuration  will corresponds to that of the heat bath. We thus
solve the equation of motion of sigma  model  fields  in  contact
with  heat bath which we represent as a white noise source. To be
consistent with fluctuation-dissipation theorem, we also  include
a  friction  term  in  the equation of motion. We thus propose to
study following Langevin equation for O(4) fields,

\begin{equation}
[\Box +\eta\partial_t +\lambda (\Phi^2-f^2_\pi)]\Phi=H n_\sigma + \zeta
\label{1}
\end{equation}

\noindent  where $\eta$ is the friction. The heat bath $\zeta$ is
represented by  a  white  noise  source  with  zero  average  and
correlation as demanded by fluctuation-dissipation theorem,

\begin{mathletters}
\begin{eqnarray}
<\zeta(t,x,y,z)> =&&0\\
\int <\zeta_a(t_1,x_1,y_1,z_1)\zeta_b(t_2,x_2,y_2,z_2)> d^4x
=&& 2 T \eta \delta_{ab}
\end{eqnarray}
\label{1a}
\end{mathletters}

Recently  it  was shown that in $\phi^4$ theory hard modes can be
integrated out on the two loop level leading to  dissipation  and
noise in the quasi classical limit for the propagation of the long
wavelength fields \cite{gr97}. This also justify our approach.

If the fields are allowed to evolve in contact with the heat bath
at  temperature  T  for a long time, they will be equilibrated at
that temperature.  Above  a  certain  critical  temperature,  the
fields  will  undergo symmetry restoring phase transition, if the
model  allows  for  such  a  transition.  For  sigma  model,  the
condensate  value  of the sigma field provide a convenient way to
probe  the  phase  transition.  In  the  symmetric   phase,   the
condensate  is  zero,  while  in  the symmetry broken phase it is
non-zero.

In the present paper we will consider linear sigma model without
the symmetry breaking term. It is well known that there will not be
an exact phase transition if the symmetry breaking term is included.
We  solve  the  eq.\ref{1}  on  a  $32^3$  lattice  with periodic
boundary   condition.   Solving   eqn.\ref{1}   require   initial
conditions  ($\phi$  and $\dot{\phi}$).
 We distribute the initial
fields according to a random Gaussian with

\begin{mathletters}
\begin{eqnarray}
<\sigma>=&&f(r)f_\pi \\
<\pi_i>=&&0 \\
<\sigma^2>-<\sigma>^2 = <\pi_i^2>-<\pi_i>^2=   && v^2/4 f(r)\\
< \dot{\sigma}>=&& <\dot{\pi_i}>=0\\
<\dot{\sigma}^2>=<\dot{\pi}>^2=&& v^2
\end{eqnarray}
\label{2}
\end{mathletters}

The interpolation function

\begin{equation}
f(r)=[1+exp(r-r_0)/\Gamma)]^{-1}
\end{equation}

\noindent  separates  the  central  region  from  the rest of the
system.
We have used $r_0=11a$ where $a$ is  the  lattice  spacing  and
$\Gamma$=0.5  fm.
The initial field configuration corresponds to zero
temperature, $<\sigma> \sim f_\pi$, 
$<\roarrow{\pi}> \sim 0$. 
Initial zero temperature fields will thermalise
to the temperature of the heat bath if kept in contact for sufficient time.
The other parameter of  the  model
is  the  friction  ($\eta$). In the present paper, we use $\eta =
\eta_\pi +\eta_\sigma$ and for $\eta_\pi$ and  $\eta_\sigma$  use
values  as  calculated by Rischke \cite{ri98} but  its
precise value is not of importance here, as we  are  looking  for
fields  at  equilibrium. Friction determines the rate of approach
to equilibrium. This aspect of equilibration was verified.

We  define  volume  averaged  sigma  condensate $<\sigma>$ as the
order parameter,

\begin{equation}
<\sigma> = 1/V \int d^3x \sigma
\end{equation}

In  fig.1,  we  have  shown  the  equilibrium  value of the order
parameter as  a  function  of  the  temperature.  Results  for  4
different lattice spacing, a=0.5, 1.0, 2.0 and 2.5 are shown. For
all  the  lattice  spacings,  the  order parameter decreases from
$\sim f_\pi$ at very  low  temperature  to  exact  zero  at  some
critical  temperature  ($T_c$)  and  then  remain  so  beyond the
critical temperature. Phase transition in the model  is  evident.
However it is also evident that phase transition temperature (the
temperature   at   which  $\sigma$  condensate  vanishes)  depend
strongly on the lattice spacing used.  It  goes  to  smaller  and
smaller  values  as the lattice spacing is reduced. It seems that
in the continuum limit, $T_c \rightarrow 0$.

\begin{figure}
\centerline{\psfig{figure=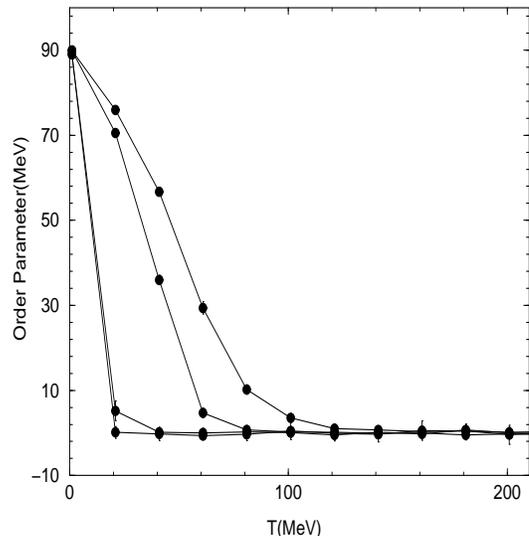,height=7cm,width=7cm}}
\caption{Order parameter as a function of temperature for four different
lattice spacing, a=0.5, 1.0, 2.0 and 2.5. The outer ones are for higher lattice
spacing}
\end{figure}

The  present  result  indicate  that classical sigma model fields
donot  show  correct  equilibrium  behaviour.  Phase   transition
temperature  depend  strongly  on the lattice spacing, and in the
continuum limit, it goes to zero or atleast to a very low temperature.
Strong dependence of the the transition temperature on lattice spacings
indicate that the simulation studies of disoriented chiral condensate using
classical sigma model fields are highly questionable. The results obtained
from those simulations can not be believed to represent physical systems.
 

The author would like to thank 
X.-N. Wang,  J.  Randrup  and V. Koch for useful
discussions.
He also  acknowledges the kind hospitality of Lawrence
Berkeley Laboratory, where part of the work was done.

\end{document}